\documentclass[a4paper,11pt]{article}
\usepackage{pos}
\usepackage{braket}
\usepackage{epsfig,color}

\usepackage{graphicx} 
\usepackage{braket}
\usepackage{amsmath}
\usepackage{tensor}
\usepackage{physics}
\usepackage{hyperref}
\usepackage{stackengine}
\usepackage{comment}
\usepackage{url}
\usepackage{slashed}
\usepackage{simpler-wick}
\usepackage[utf8]{inputenc}
\usepackage{mathtools}
\usepackage{tikz}
\usetikzlibrary{decorations.markings}

\usepackage{rotating}

\newcommand{\MSbar}{\overline{\text{MS}}}

\ShortTitle{Agentic AI for QCD perturbation theory}
\title{Agentic AI for QCD perturbation theory or: how I learned to stop worrying and love the bot}

\author*[a]{Erik Lundstrum}

\affiliation[a]{Physics Department, Columbia University,\\
  New York, NY 10027, USA}

\emailAdd{ell2156@columbia.edu}
\abstract{
    AI coding agents are extremely effective at performing calculations that can be broken down into modular steps, each with verifiable checks of correctness. Perturbative QCD calculations satisfy both of these properties and we find that they are well-suited to solution by agentic AI systems. We discuss our methods for instructing and verifying the work of agents in one- and two-loop perturbative calculations, with a focus on pre-empting common mistakes. We find that once an agent has successfully completed one perturbative calculation, its instructions transfer very well to other similar calculations which can then be completed and verified in a matter of hours with little effort from the user.}

\FullConference{The 43rd International Symposium on Lattice Field Theory (Lattice 2026)\\
July 26 to August 1, 2026\\
University of Maryland, College Park, USA\\}

\begin{document}
\maketitle

\section{Introduction}
AI coding agents have revolutionized our programming ability. For many simple tasks for which little verification is needed we can often assume the output is correct. For difficult physics calculations which require thousands of lines of code, the situation is more delicate. One possible verification strategy is of course to read through every line of code the agent produces. However, the real utility of these agents is to free ourselves from having to personally interact with large code bases. 

In this paper, we describe our agentic AI strategy for performing and verifying complicated perturbative QCD calculations. We find that an effective strategy is to make the calculation modular by breaking it into small building blocks, each of which is of appropriate size for an agent to complete. In our case, a small building block is a short, self-contained bit of Mathematica code. These building blocks should be cleverly chosen so that each is independent and can be checked for correctness on its own. We can then put these blocks together to build up verified solution chains, the final results of which can be trusted as correct after we apply a series of global checks of correctness. In the case of perturbative QCD there are many internal consistency checks which serve to verify a calculation. Besides giving us a solution we can trust, this strategy also frees us from having to read much of the actual code the agent is writing; as long as the expected output is confirmed against an input, we can be sure that the code has been correctly implemented. 

The calculations we focus on here are the perturbative QCD calculations required to renormalize lattice QCD simulations of weak decays, represented by effective four-quark operators. We have so far completed three perturbative calculations of the RI-$\MSbar$ matching factors. The three calculations are: 1) The one-loop matching factors required for the $\Delta S = 1$ four quark operators with four active quark flavours, 2) the two-loop matching factors required to renormalize the long-distance contribution to the CP-violating parameter $\epsilon_K$ in which the $\Delta S = 2$ process proceeds via two $\Delta S = 1$ operators and 3) the one-loop matching factors required for several $\Delta I = 2$ operators relevant for neutrinoless double beta decay. The final calculation was completed in a matter of hours by transferring the agent which had been instructed for the first calculation.

The outline of this paper is as follows. In Section \ref{sec:QCD-pert} we briefly describe the perturbative QCD calculations we have performed. In Section \ref{sec:AI-strat}, we describe our method for instructing the agents to complete these calculations and the various checks we can use to confirm correctness. In Section \ref{sec:transferring} we discuss how well agents that have been instructed to solve one problem perform when given similar problems. Finally, we give our conclusions in Section \ref{sec:conclusion}.

\section{QCD corrections to effective weak operators}
\label{sec:QCD-pert}
\begin{figure}[h]
  \centering
  \includegraphics[scale=0.3]{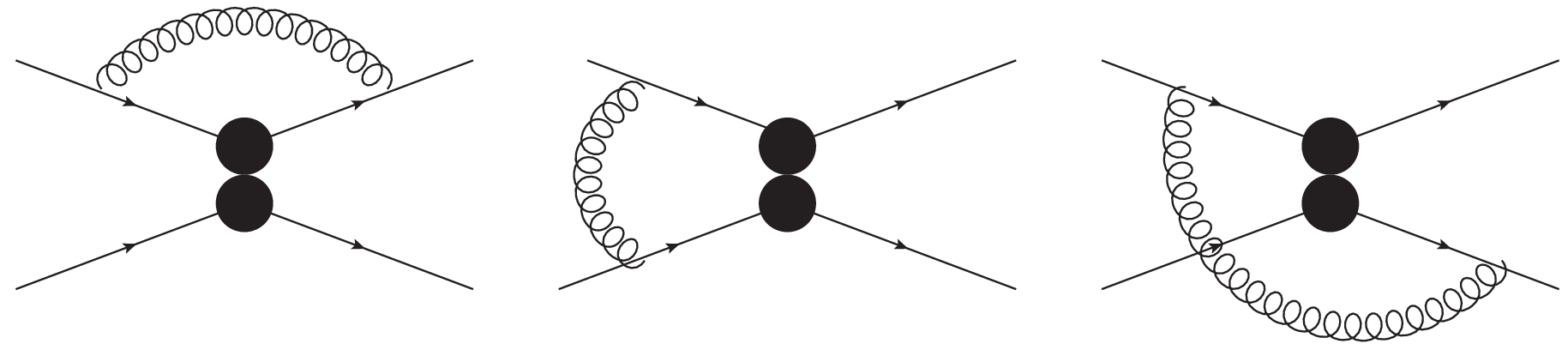} 
    \caption{One-loop QCD corrections to the $\Delta S = 1$ four-quark operators.}
    \label{fig:QCD-corrections-four-quark}
\end{figure}
\begin{figure}[h]
  \centering
  \includegraphics[scale=0.3]{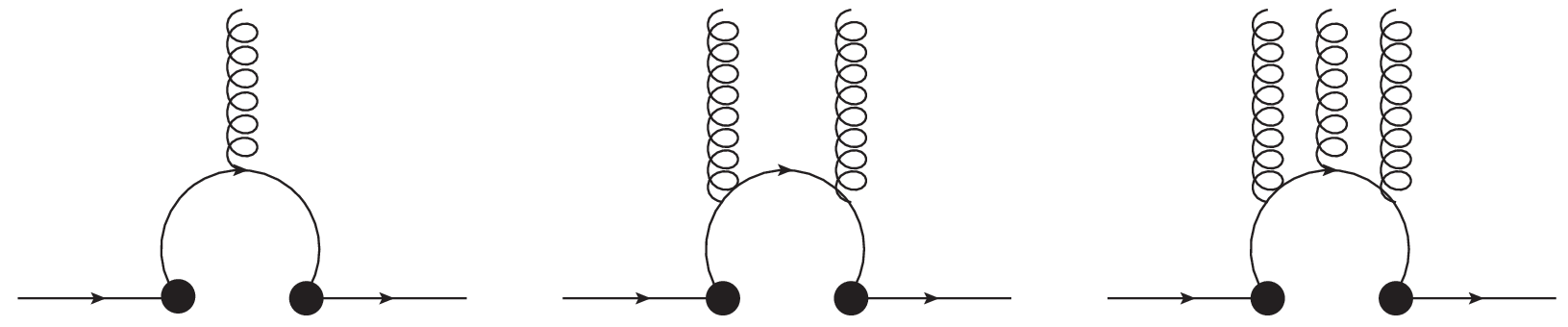} 
   \caption{One loop diagrams responsible for the mixing of the four-quark operators with penguin operators. In addition to these diagrams one must consider those in which gluons are emitted from closed fermion loops.}
    \label{fig:penguin-diag}
\end{figure}

In this section we briefly describe the perturbative calculations required to renormalize the matrix elements of weak operators determined in lattice QCD simulations. We use our first calculation of the one-loop renormalization factors of the $\Delta S = 1$ operators to motivate these calculations. Following this, we discuss the details of our two-loop calculation relevant for $\epsilon_K$. The details of the third calculation are deferred to Section ~\ref{sec:transferring}.

\subsection{$\Delta S = 1$ operators}
\label{sec:QCD-pert-deltaS-1}

The asymptotic freedom of QCD means that at high energies the strong coupling $\alpha_s$ is small enough that we can reliably use QCD + electroweak perturbation theory. At the energy scales relevant to hadronic physics the coupling becomes large, resulting in a strongly-coupled theory. In order to simulate hadronic processes, we rely on the operator product expansion (OPE). This allows us to write the weak interaction Hamiltonian for the $\Delta S = 1$ process as
\begin{equation}
    H_W^{\Delta S = 1} = \frac{G_F}{\sqrt{2}}\sum_{i} C_i(\mu) Q_i(\mu).
\label{eq:H-eff-dS-1}
\end{equation}
Here, the Wilson coefficients $C_i$ capture the short-distance physics and are calculated perturbatively in QCD + electroweak perturbation theory \cite{Buchalla:1995vs}. The operators $Q_i$ are four-quark operators that effectively realize the weak interactions on the lattice at energy scales far below the $W$ boson mass where the charged current exchange appears to be point-like. Example operators include the $\Delta S = 1$ current-current operators
\begin{equation}
    Q^{q' \overline{q}}_1 = (\bar{s}_a q'_b)_{V-A} (\bar{q}_b d_a)_{V-A}, \qquad
    Q^{q' \overline{q}}_2 = (\bar{s}_a q'_a)_{V-A} (\bar{q}_b d_b)_{V-A},
\label{eq:current-current-operators}
\end{equation}
where $q,q' \in \{u,c\}$ and $(\overline{q}q)_{V \pm A} = \overline{q} \gamma_\mu (1 \pm \gamma_5) q$ in the quark bilinears. In our calculations we also consider QCD and electroweak penguin operators but we omit these here in the interest of brevity.

The problem is that perturbative calculations are typically performed with dimensional regularization in the $\MSbar$ scheme, while lattice QCD calculations are lattice-regularized. In order to combine the two quantities, we use the regularization-independent renormalization scheme as an intermediate scheme ~\cite{Martinelli:1994ty}. To do so, we need to calculate conversion factors $Z^{lat \rightarrow RI}_{ij}$ and $Z^{RI \rightarrow \MSbar}_{ij}$ which are inserted between the Wilson coefficients and operators in Eq.~\eqref{eq:H-eff-dS-1}. The purpose of our calculations is to determine the factor $Z^{RI \rightarrow \MSbar}_{ij}$ which is calculated in perturbation theory. For a detailed discussion of the method for calculating the RI - $\MSbar$ matching factors for four-quark operators see Ref.~\cite{Lehner:2011fz}. 

To determine the RI-$\MSbar$ matching factors we require that suitable Green's functions computed between off-shell quark and gluon external states, in a fixed gauge, coincide with their tree-level value. To impose this condition we calculate the $\MSbar$-renormalized Green's functions arising from the diagrams in Figs. \ref{fig:QCD-corrections-four-quark} and \ref{fig:penguin-diag} with a covariant gauge-fixing parameter $\xi$, to which we apply an appropriate set of projectors onto spin, colour and flavour indices. To limit the scope of this paper we focus on evaluating the $\MSbar$-renormalized Green's functions. We also omit various subtleties related to dimensional regularization and evanescent operators. Details can be found in the references given.

\subsection{Bi-local $\Delta S = 2$ operators}
\label{sec:QCD-pert-epsK}
\begin{figure}[h]
  \centering
  \includegraphics[scale=0.3]{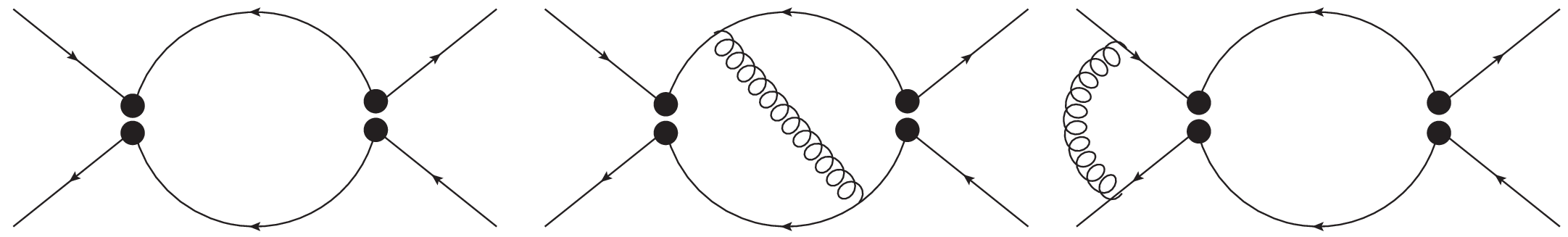} 
   \caption{One- and two-loop diagrams appearing in the bi-local operator contribution to the $\Delta S = 2$ process. The four-fermion vertices are $\Delta S = 1$ operators such as those in Eq.~\eqref{eq:current-current-operators}.}
    \label{fig:bi-local}
\end{figure}
Our second calculation is a similar set of matching factors required to renormalize the lattice QCD calculation of the long-distance contribution to $\epsilon_K$ \cite{Bai:2023lkr, Huo:2025bhq}. In this case, the $\Delta S = 2$ process proceeds via two $\Delta S = 1$ operators. Example diagrams contributing to this process are shown in Fig. \ref{fig:bi-local}. In this case the $O(\alpha_s)$ contributions first arise at two loops. The effective interaction Hamiltonian in this case is
\begin{equation}
H^{\Delta S = 2}_W
=
\frac{1}{2}T\left\{
\int d^4x\,H_W^{\Delta S=1}(x,\mu)H_W^{\Delta S=1}(0,\mu)
\right\}
+
C_{\Delta S = 2}
Q_{LL}(0),
\label{eq:H-dS-2-eff}
\end{equation}
where $H^{\Delta S = 1}_W$ is the interaction Hamiltonian given in Eq.~\eqref{eq:H-eff-dS-1} and $Q_{LL}$ is the local $\Delta S = 2$ operator
\begin{equation}
     Q_{LL}^{\Delta S = 2} = (\overline{s}_a d_a)_{V-A} (\overline{s}_b d_b)_{V-A}.
\label{eq:dS-2}
\end{equation}
In this case the collision of the two $\Delta S = 1$ operators in the integral $\int d^4 x$ results in a logarithmic divergence proportional to the $\Delta S = 2$ operator $Q_{LL}$. Therefore, the Wilson coefficient $C_{\Delta S = 2}$ must also contain a counter-term to cancel the divergence arising as $x \rightarrow 0$. The effect of the RI-$\MSbar$ matching factor will be to shift the finite part of the Wilson coefficient $C_{\Delta S = 2}$ used in the lattice QCD calculation. The magnitude of the shift is proportional to $\sum_{i,j}C_i C_j Y_{ij}$, where $C$ are the $\Delta S = 1$ Wilson coefficients and $Y_{ij}$ is the result of applying the RI projector to the $\MSbar$ amplitude with the $\Delta S = 1$ operators $Q_i$ and $Q_j$ inserted at each vertex in Fig. \ref{fig:bi-local}.

\section{Agentic AI strategy}
\label{sec:AI-strat}
This section describes the main point of this talk which is the agentic AI strategy we used to complete the above calculations and the properties of QCD perturbation theory which allow us to validate the results. We begin with a concise summary of our strategy, after which we illustrate the main  ideas with a detailed description of our strategies for our one- and two-loop calculations.

Our goal is to break up the overall problem into many small, short horizon tasks which can be independently verified. A particularly effective way to do this is to offload as much of the work as possible to verified tools like Mathematica, FeynCalc \cite{Shtabovenko:2023idz}, LiteRed \cite{Lee:2012cn} etc. Having verified each successive step, we build up a verified solution chain which can then be applied to all the relevant diagrams. We find that the scaling of human effort with the number of diagrams is remarkably flat since the agent can apply the verified code to hundreds of diagrams with few mistakes. After verifying the functionality on a diagram-by-diagram basis, we apply to the final result a number of global checks of correctness related to the divergence structure and gauge symmetry. Subsets of results can also often be compared to existing results in the literature. Our observation is that the agents (in our case the models Claude Opus and Fable) make very few of the common human typo or algebraic mistakes, with mistakes instead arising due to the agent making incorrect assumptions after the user has given an insufficiently detailed prompt. Of course the user must understand the problem well enough to make sufficiently detailed prompts.

We note that in our experience a very common source of issues was Claude attempting to write its own functionality from scratch, despite being instructed at the outset to use as much functionality as possible from verified tools such as FeynCalc or LiteRed. The issue seemed to be that Claude was unaware of or unprepared to use existing functionality and had to be reminded regularly to search for functionality in an existing tool before attempting to create its own.

\subsection{One-loop $\Delta S = 1$ calculation}
\label{sec:one-loop}
In this subsection we describe the above strategy as we applied it to our one-loop calculation of the matching factors described in Sec. \ref{sec:QCD-pert-deltaS-1}. A paper detailing this calculation will appear shortly \cite{Lehner-Lundstrum}. The relevant diagrams are shown in Figs.\ref{fig:QCD-corrections-four-quark} and \ref{fig:penguin-diag}. We work with four flavours of massless quark and in a covariant gauge specified by the parameter $\xi$ where $\xi=0$ corresponds to Landau gauge and $\xi = 1$ to Feynman gauge. There are only 12 diagrams required for this calculation, and we enumerated them in hand-drawn images with the external quark flavours and momenta labelled and gave them as images to Claude. Our solution chain for this calculation is as follows, each of which will be discussed in more detail. First, identify and evaluate all required tensor integrals and save the results. Second, for each diagram, identify the Dirac structures in the numerator and combine with the cached integral results. Third, project the many resulting spinor structures onto the basis defined in Ref. \cite{Lehner:2011fz}. Fourth, combine diagrams as required to find the one-loop correction to each operator $Q_1 - Q_{10}$. Finally, we apply global checks of correctness.

To illustrate our strategy we focus on one example loop integral that appears in the calculation
\begin{equation}
        \int \frac{d^D k}{(2\pi)^D} \frac{1}{k^2}\Big( \eta^{\mu\nu} - (1 - \xi) \frac{k^\mu k^\nu}{k^2}\Big) \gamma_\mu \frac{(\slashed{p}_1 + \slashed{k})}{(p_1 + k)^2} \gamma^\alpha P_L \frac{(\slashed{p}_2 + \slashed{k})}{(p_2 + k)^2} \gamma_\nu.
\end{equation}
Here, $p_1$ and $p_2$ are external momenta carried by the quarks and $P_L = (1 - \gamma_5)$. The free $\alpha$ index contracts with the other quark bilinear. The two operators in Eq.~\eqref{eq:current-current-operators} differ only by colour indices and therefore we can use the same integrals for the radiative corrections to both operators. The first step is to remove all the Dirac structures and have the agent create a verified step that just solves the tensor integral. To do this we simply have the agent create Mathematica code that enumerates the tensor integrals with numerator structures $\{1, k^\mu, k^\mu k^\nu,  k^\mu k^\nu k^\rho, k^\mu k^\nu k^\rho k^\sigma \}$ and the appropriate additional factors of $1/k^2$ from the gauge-dependent part of the gluon propagator. We then use FeynCalc to evaluate the integrals in terms of a set of master scalar integrals. This step is easy to verify by simply reading the Mathematica code that Claude produces. We save the verified integral results for use in the next step.

The next step is to read back in the verified tensor integral results and combine them with the outstanding Lorentz indices on the Dirac structures that were isolated in the previous step. Doing so leaves us with a large number of Dirac structures such as
\begin{equation}
\renewcommand{\arraystretch}{2.2}
\begin{array}{@{}c@{\qquad}c@{\qquad}c@{}}
    \Big( \overline{s} \gamma^\mu P_L u \Big) \Big(\overline{u} \gamma_\mu P_L d\Big),
  & \Big( \overline{s} \slashed{p}_1 \slashed{p}_2 \gamma^\mu P_L u \Big) \Big(\overline{u} \gamma_\mu P_L d\Big),
  & \Big( \overline{s} \slashed{p}_1 P_L u \Big) \Big(\overline{u} \slashed{p}_1 P_L d\Big), \\
    \Big( \overline{s} \slashed{p}_1 \gamma^\nu \gamma^\mu P_L u \Big), \Big(\overline{u} \gamma_\mu \slashed{p}_2 \gamma_\nu P_L d\Big)
  & \Big( \overline{s} \slashed{p}_1 P_L u \Big) \Big(\overline{u} \slashed{p}_2 P_L d\Big),
  & \Big( \overline{s} \gamma^\sigma \gamma^\nu \gamma^\mu P_L u \Big) \Big(\overline{u} \gamma_\mu \gamma_\sigma \gamma_\nu P_L d\Big).
\end{array}
\end{equation}
We want to express all of these terms in a standard basis used in Ref. \cite{Lehner:2011fz}. To do this, we have the agent write Mathematica code to project arbitrary Dirac structures into the  basis we chose. To verify this step, we perform several of the reductions with pen and paper using gamma matrix identities, then require that the agent reproduce our results. Once verified, we have a projection code that can be applied to all spinor structures that appear in the calculation.

Next, we want to combine the diagrams with the correct colour factors. To determine the relevant colour factors we use the following property of the SU($N_c$) generators $T^a$
\begin{equation}
    \sum_{a = 1}^{N_c^2 - 1} T^a_{ij} T^a_{kl} = \frac{1}{2} \left(\delta_{il} \delta_{jk} - \frac{1}{N_c} \delta_{ij}\delta_{kl} \right).
\end{equation}
We instructed Claude to use the above property to find the colour factors relevant for each diagram and operator in Eq.~\eqref{eq:current-current-operators} and verified its calculation manually. 

Finally, we apply global checks of correctness to our results. We use the following checks:
\begin{enumerate}
    \item Identify the gauge-dependent $1/\epsilon$ pole corresponding to the quark wavefunction renormalization factor $Z_q^{-2}$ in each of the final amplitudes.
    \item Confirm that any structure containing external momenta $p_{1,2}$ does not have a $1/\epsilon$ pole.
    \item Check that after applying $Z_q^{-2}$ the remaining poles are gauge-independent.
    \item Finally, we can check the values of our poles against those appearing in the literature \cite{Buchalla:1995vs, Lehner:2011fz}, adjusting for the number of active quark flavours.
\end{enumerate}
We found that each of these checks revealed issues with our calculation. These issues were all related to Claude making incorrect assumptions in intermediate steps when not given sufficiently strict instructions. Examples include inconsistent use of $D$- vs. four-dimensional Dirac algebras, different steps using different treatments of $\gamma_5$ in dimensional regularization and inappropriately applying Fierz transformations before being instructed to do so. As a final check of correctness, we compared our results against those of a collaborator who performed the same calculation without the use of Claude. Once the above checks were satisfied, we found complete agreement on the first comparison of our results.

\subsection{Two-loop $\Delta S = 2$ calculation}
\label{sec:two-loop}
In this section we discuss the more difficult two-loop calculation described in Section \ref{sec:QCD-pert-epsK}. This calculation is described in detail in Ref.~\cite{Lundstrum:2026lvb}. Here we focus on the agentic AI strategy and our verification process.  To complete this calculation, we adapt the strategy for calculating the RI-$\MSbar$ matching factors that was used in Ref. \cite{Gorbahn:2024qpe} for the $\Delta S = 2$ operator in Eq.~\eqref{eq:dS-2}. The idea is to take advantage of the fact that only a subset of the spinor structures which appear in the $\MSbar$-renormalized amplitudes actually contribute to the RI-$\MSbar$ matching factors in the relevant scheme. We therefore define a set of projectors onto the relevant spinor structures, which we apply to the un-evaluated loop integrands. This removes all spinor and Lorentz indices, resulting in a cheaper set of scalar two-loop integrals that must be evaluated. We verify the code which projects onto the relevant spinor structures by providing a set of structures with known projections. 

First, we enumerated all of the diagrams contributing to the bi-local operator at two loops. Some example diagrams are shown in Fig. \ref{fig:bi-local}. To classify the diagrams, we first provided to Claude hand-drawn images of one diagram from each of the unique topologies with the external momenta labelled. For example, we would provide the right-most diagram in Fig. \ref{fig:bi-local} and tell it to also prepare code for the diagram of the same class where the gluon line is on the right side instead of the left. We verified in a few of the cases that Claude was able to take a hand-drawn image of a diagram and apply the Feynman rules to produce Mathematica code for the integrand. Throughout the calculation, we encountered no instances of Claude making mistakes in applying the Feynman rules. We later switched to a more rigorous strategy of using QGRAF \cite{qgraf} to enumerate all of the relevant diagrams and having Claude write Mathematica code based on the output. We found that the results agreed with those from our original hand-drawn method. 

Our next step was to instruct Claude to apply the verified projection code to each of the diagrams. Doing so, we are left with one scalar integral for each projector and diagram. Next, we instructed Claude to feed these scalar integrals to integration by parts (IBP) reducer packages to express our results in terms of a set of master scalar integrals. For the simpler classes of two-loop diagrams we used LiteRed \cite{Lee:2012cn} to perform the IBP, but for the more complicated diagrams we switched to the more performant FIRE6 \cite{Fire6}. We also used TARCER \cite{Tarcer} for a set of integrals that have closed analytic solutions. Finally, the master scalar integrals which had no known analytic solution were evaluated numerically by having Claude use PySecDec \cite{pysecdec}. Once we verified the diagram enumeration and projection code, the remaining steps given above were just a matter of Claude writing a large amount of code that essentially copied the output of one step to the input to another. We found that Claude excelled in simple tasks like these as there is not much opportunity for it to make incorrect assumptions in intermediate steps. Having gained more experience with the common sources of Claude's mistakes we performed fewer tests on the intermediate steps and relied more upon the global checks of correctness discussed below. 

Next, we discuss the global checks we used to verify our results. The divergence structure of higher-order perturbative calculations is tightly constrained by gauge invariance, locality and the renormalization group (RG), giving us several verification checks. Our calculation was performed with the covariant gauge fixing parameter $\xi$ left symbolic. The first check of our results is that the poles should be gauge-independent after the poles corresponding to the quark wavefunction renormalization have been removed. The second check we can apply is the cancellation of non-local divergences like $\ln(p^2)/\epsilon$ after the insertion of one-loop counterterms required to cancel sub-divergences. As a final check of the divergence structure, the RG constrains the coefficients of the $\alpha_s^2/\epsilon^2$ poles from the coefficients of the one-loop divergences, and therefore we can check that all of our $\alpha_s^2/\epsilon^2$ poles agree with their values predicted by the one-loop calculations. Having performed these internal consistency checks, we confirmed our results for the genuine two-loop poles $\alpha_s^2 / \epsilon$ by comparing to results that have appeared previously in the literature \cite{Brod:2010mj}. We also applied internal consistency checks for the finite parts of our amplitudes. One check was of the scaling of our finite amplitudes with the charm quark mass. Since the bi-local operator is dimension eight, we should find at most $m^2_c$ scaling. This check revealed an issue due to Claude using two different notations for $m_c$ in different sets of diagrams. We also checked the exchange symmetry of the s- and t-channel diagrams under swapping down quark operators and external momenta. As a final check of our results, we also verified the finite part of our results in the limit of vanishing external momentum to those which were previously calculated with zero external momentum in Ref. \cite{Brod:2010mj}. We found that all of these finite parts agreed on the first comparison after applying the above checks. 

\section{Transferring pre-instructed agents}
\label{sec:transferring}
In this section we note the general applicability of the agents whose instruction we discussed in the previous sections. As an experiment, we compute the one-loop RI-$\MSbar$ matching factors for several operators contributing to neutrinoless double beta decay. Two examples are
\begin{equation}
    O_{1}^\mu = (\bar{u}_a\gamma^\mu P_L d_a)(\bar{u}_b P_L d_b), \qquad O_{2}^\mu = (\bar{u}_a\gamma^\mu P_L d_b)(\bar{u}_b P_L d_a).
\end{equation}
To do this, we made a fresh agent and gave it access to the directory containing the CLAUDE.md file and code used to perform the $\Delta S = 1$ calculation discussed in Sec. \ref{sec:one-loop}. These calculations share the same diagrams shown in Fig.\ref{fig:QCD-corrections-four-quark}, but the diagrams in Fig. \ref{fig:penguin-diag} do not contribute. We instructed Claude to use the exact same strategy and steps, replacing only the spinor structures that appear in the calculation. After this, we applied the same checks as were described in Sec. \ref{sec:one-loop} and found that on the first attempt the results for our divergences agreed with those calculated previously in Ref. \cite{Cirigliano:2018yza}. The entire process from identifying the relevant operators to obtaining the finite amplitudes and verifying the pole coefficients took only a few hours.

\section{Conclusion and outlook}
\label{sec:conclusion}

In this talk we discussed our experience using agentic AI to complete  one- and two-loop perturbative QCD calculations required to renormalize lattice QCD simulations of weak decays. We find that perturbative QCD is particularly well-suited to solution by agentic AI because of the large number of internal consistency checks and software packages such as FeynCalc with which the agent can interface. We also found that once an agent has been instructed to complete one calculation its instructions generalize very well to similar calculations, allowing them to be completed in a matter of hours. We conclude that one- and two-loop perturbative QCD calculations are now very accessible for the average lattice QCD practitioner. We also expect that the strategies discussed here will generalize very well to higher loop orders. It should be possible in the near future to perform perturbative QCD calculations to higher orders than has previously been feasible.

\bibliographystyle{JHEP-2}
\bibliography{ref}

\end{document}